\documentclass[times,doublespace]{simauth}

\usepackage{moreverb}

\usepackage[dvips, colorlinks,bookmarksopen,bookmarksnumbered,citecolor=red,urlcolor=red]{hyperref}

\newcommand\BibTeX{{\rmfamily B\kern-.05em \textsc{i\kern-.025em b}\kern-.08em
T\kern-.1667em\lower.7ex\hbox{E}\kern-.125emX}}

\def\volumeyear{2010}
\begin{document}

\title{Nonparametric ROC Summary Statistics for Correlated Diagnostic Marker Data}
   \author{Liansheng Larry Tang$ ^{1 }$,  Aiyi Liu${^{2}}$, Zhen Chen${^{2}}$, Enrique F. Schisterman${^{2}}$, Bo Zhang$^{3}$, and Zhuang Miao$ ^{1 }$  }
\address{$^{1}$ Department of Statistics,  George Mason University, Fairfax, VA 22030, USA\\
$^{2}$ Division of Epidemiology, Statistics and Prevention Research, National Institute of Child Health and Human Development, Rockville, Maryland 20852, USA\\
$^{3}$ School of Biological and Population Health Sciences,
College of Public Health and Human Sciences,
Oregon State University,
Corvallis, 97331, USA}
\corraddr{ltang1@gmu.edu, Department of Statistics,  George Mason University, Fairfax, VA 22030, USA}

\begin{abstract}
 We propose efficient nonparametric   statistics  to compare medical imaging modalities in multi-reader multi-test data and to compare markers in longitudinal ROC data. The proposed methods are based on the weighted area  under the ROC curve which includes the area under the curve and the partial area under the curve as special cases. The methods maximize  the local power for detecting the difference between imaging modalities.   The  asymptotic results of the proposed methods    are developed under a complex  correlation structure.  Our simulation studies show that  the proposed statistics result in  much better powers than existing statistics.  We applied the proposed   statistics  to an endometriosis
diagnosis study.
\end{abstract}


\maketitle
 \section{Introduction}
In medical imaging studies,  one is concerned about whether  a newly developed imaging modality  is more accurate than traditional modalities to  correctly  discriminate   a subject with abnormal lesions from a subject without such lesions.
 Imaging modalities  are considered as an example  of diagnostic markers, which are used to distinguish a subject with a particular condition (``the diseased'') from a subject without the condition (``the non-diseased'').
For
diagnostic markers  that generate  binary test results, their accuracy can be
summarized in terms of sensitivity (probability of
identifying a diseased subject when the disease truly exists) and specificity (probability of
correctly ruling out a non-diseased subject when the disease is truly absent). For diagnostic markers
that
 generate   discrete or continuous test results, the receiver operating characteristic (ROC) curve
is a standard statistical tool to describe and compare the accuracy of  markers \cite{zhou:02}. The ROC curve   combines  all possible pairs of sensitivities
  and 1$-$specificities      from different decision thresholds
and thus    describes the accuracy of markers   apart from
decision thresholds.

    For correlated results from two diagnostic markers, parametric and nonparametric  methods have been proposed  to  compare   ROC summary measures. Parametric methods   for  the area  under the curve (AUC) assume  distributions (e.g. negative exponential,  normal,
 lognormal, gamma) on marker measurements \cite{zou:01, molodianovitch:06}. These methods may not perform well if the parametric assumptions are invalid.  The semiparametric ROC estimation based on the logistic regression is proposed by \cite{Copas:02}. As an alternative, nonparametric methods do not require distribution assumptions and are robust to model misidentification.  Nonparametric methods to estimate and compare two AUCs have been proposed by \cite{delong:88}, \cite{obuchowski:97}, and others. These methods are based on results for U-statistics because an empirical AUC statistic is essentially a Wilcoxon rank sum statistic \cite{bamber:75}.   However, if  two ROC curves intersect, their AUCs may be equal and do not provide valid information for the comparison. Moreover, summarizing the entire ROC curve may  include irrelevant information about the marker's accuracy when one is only interested in some range of   specificities. For example, acceptable specificities are high for early cancer detection tests.     The partial area under the curve  (pAUC), which summarizes part of the ROC curve  in the range of desired specificities, may be  a better alternative. Nonparametric methods   to compare pAUCs are proposed by \cite{zhang:02}. \textbf{{{Utilizing the pAUCs is particularly important in comparing  markers which are developed to screen a large population for certain diseases, for example, breast cancer \cite{baker:01}.  A lower specificity for a large population leads to many more falsely classified non-diseased subjects who may have to undergo a more invasive test subsequently. It is thus desired to compare screening markers at a higher range of specificities. }}}

 {
\textbf{In this paper we propose efficient nonparametric ROC statistics to analyze    multi-reader multi-test ROC data and to nonparametrically summarize correlated longitudinal ROC data. The proposed method not only includes many nonparametric ROC summary measures as special cases, but also  maximizes  the local power for detecting the difference between markers.}}
%In addition, the new statistics   include   generalized Wilcoxon statistics for correlated data discussed by \cite{rosner:03}.
%The theoretical   results of  the new statistics are  naturally  equivalent to those of the generalized Wilcoxon statistics under specific conditions.
  The rest of the article is organized as follows. In  Section  \ref{sec:notation}   we   introduce the new statistics for multi-reader multi-test ROC data and longitudinal ROC data, and discuss the  equivalence  between our statistics and the generalized Wilcoxon statistics under specific assumptions. Section \ref{sec:deltas} gives the   variance expressions for the proposed statistics. Section \ref{sec:sim} reports simulation results to illustrate the small sample performance of  the proposed ROC statistics and their theoretical variances.  Section \ref{sec:example1} applies the proposed method to a real example on the diagnosis of endometriosis.  Section  \ref{sec:dis} gives some discusion.

\section{Methods}\label{sec:notation}
\subsection{Definition of nonparametric ROC summary statistics}
We first define some notations.   Suppose test result $X_{\ell i p }$ of marker $\ell$ is from the   $p $th abnormal location  in the diseased subject $i$, where $\ell=1,...,L$, $p  =0,1,...,m _{\ell i}$, and  $i=1,...M$.  Test result  $Y_{\ell j q }$ of  marker $\ell$ is from  the $q $th normal location  in the non-diseased subject $j$, where $\ell=1,...,L$, $q =0,1,...,n _{\ell j}$, and  $j=1,...J$. Here the total number of subjects is $N=M+J$.  The joint pairwise cumulative  function of $(X_{\ell_1 i p_1},X_{\ell_2 i p_2})$ is taken to be $S_{D,\ell_1, \ell_2}(x_1,x_2)$, $p_1,p_2=1,...,m_{\ell i}$,  with marginal survival functions $X_{\ell  i p} \sim
S_{D,\ell }(x)$. Similarly we define    $(Y_{\ell_1 j q_1}, Y_{\ell_2 j q_2}) \sim
S_{\bar D,\ell_1, \ell_2}(y_1,y_2)$, $q_1,q_2=1,...,n_{\ell i}$,  with marginal survival functions $Y_{\ell j q } \sim S_{\bar D,\ell}(y)$.
The  ROC curve
for the $\ell$th marker is then given by
 $ ROC_\ell(u) = S_{D,\ell}\left\{S_{\bar D,\ell}^{-1}(u)\right\}$, where  the false positive rate (FPR) $u$ is in $[0,1]$. The resulting $\ell$th weighted area under the curve (wAUC)  is
\begin{equation}\label{eq:omega}
\Omega_\ell =
\int_0^1 S_{D,\ell}\left\{S_{\bar D,\ell}^{-1}(u)\right\} dW(u),
\end{equation} with a probability measure $W(u)$
defined on
$u$, for $u\in [0, 1]$. Included in this class of accuracy
measures are  AUC,  pAUC  between
FPRs $u_1 $ and $ u_2 $, and the sensitivity at
a given level of FPR $  u_0 $.
$W(u)$ can also be defined as certain distribution functions, such as the beta cdf, to assign varying weight to the specificity. The detailed discussion is in \cite{li:10}.

By substituting the functions $S_{D,\ell}$ and $ S_{\bar D,\ell}$ with their respective empirical function $\hat S_{D,\ell}$ and $\hat S_{\bar D,\ell}$, the nonparametric wAUC estimator is given by $\widehat{\Omega}_\ell=
\int_0^1{ \hat {S}_{D,\ell}}\{{ \hat {S}_{\bar D,\ell}}^{-1}(u)\} dW(u).$
The empirical survival functions $\hat S_{D,\ell}$ and $\hat S_{\bar D,\ell}$ are defined  by
\begin{align} \label{eq:fsim}\hat S_{D,\ell} (x)&= \frac{1}{\sum_{i=1}^{M}m_{\ell i}}\sum_{i=1}^M\sum_{p=1}^{m_{\ell i}}I(X_{\ell i p}> x),\nonumber\\
\hat S_{\bar D,\ell} (x)&=\frac{1}{\sum_{j=1}^{J}n_{\ell j}}\sum_{j=1}^J\sum_{q=1}^{n_{\ell j}}I(Y_{\ell j q}>  x). \end{align}
Denote $  \boldsymbol{\Omega}=(\Omega_1, \Omega_2, ...,\Omega_L)$. By substituting   $\hat S_{D,\ell}$ and $\hat S_{\bar D,\ell}$ in Equation (\ref{eq:omega}), the nonparametric estimator of $\boldsymbol{\Omega}$ is given by $  \boldsymbol{\widehat{\Omega}}=(\widehat{\Omega}_1, \widehat{\Omega}_2, ...,\widehat{\Omega}_L)$.

We define  $W(u)= u$ for
$0< u < 1$ to obtain the nonparametric AUC  estimator for the $\ell$th marker as follows
\begin{equation}\label{eq:auc}
\hat \Omega_{A,\ell}=\frac{1}{\sum_{i=1}^{M}m_{\ell i}\sum_{j=1}^Jn_{\ell j} }\sum_{i=1}^M\sum_{p=1}^{m_{\ell i}}\sum_{j=1}^{J} \sum_{q=1}^{n_{\ell j}}I(X_{\ell i p}> Y_{\ell j q}).
\end{equation} The AUC statistic in (\ref{eq:auc}) takes the form of the Wilcoxon rank-sum statistic. It essentially compares the measurements of   abnormal locations with those of  normal locations.  To calculate   this statistic, we obtain every possible pair of measurements from an abnormal location and a normal location. We assign 1 if the abnormal location's measurement is larger than the normal location in the pair, and 0 otherwise. $\hat \Omega_{A,\ell}$ is then calculated by averaging   the 1's and 0's over all possible pairs.  Since the location  within each subject is viewed as the unit of sampling, the inference based on the regular Wilcoxon rank-sum statistic is not valid here.

 When $W(u) = (u-u_1)/(u_2-u_1)$ for $0 < u_1 \leq u\leq u_2 < 1$, $\widehat{\Omega}_\ell$ empirically estimates the partial AUC (pAUC), and its explicit form   is given by
\begin{equation}\label{eq:pauc}
\frac{1}{\sum_{i=1}^{M}m_{\ell i}\sum_{j=1}^Jn_{\ell j} }\sum_{i=1}^M\sum_{p=1}^{m_{\ell i}}\sum_{j=1}^{J} \sum_{q=1}^{n_{\ell j}}I(X_{\ell i p}> Y_{\ell j q}|Y_{\ell j q}\in (\hat S^{-1}_{\bar D,\ell}(u_2), \hat S^{-1}_{\bar D,\ell}(u_1)).
\end{equation}
The pAUC statistic in (\ref{eq:pauc}) uses all measurements from the abnormal locations. Since  the pAUC is specified to be in the range of   $(u_1,u_2)$,   only  measurements from the normal locations which fall in    $(\hat S^{-1}_{\bar D,\ell}(u_2), \hat S^{-1}_{\bar D,\ell}(u_1)) $ are used in (\ref{eq:pauc}). That is, we sort all measurements from the normal locations from the smallest to the largest, and obtain the order statistics $Y_{[(1-u_2)\sum_{j=1}^Jn_{\ell j}]}$ and $Y_{[(1-u_1)\sum_{j=1}^Jn_{\ell j}]}$, where $[x]$ denotes  the smallest integer greater than or equal to $x$. We then calculate the Wilcoxon rank-sum like statistic by comparing all X's with Y's which are between $Y_{[(1-u_2)\sum_{j=1}^Jn_{\ell j}]}$ and $Y_{[(1-u_1)\sum_{j=1}^Jn_{\ell j}]}$. %Thus,  (\ref{eq:pauc}) can be written as \begin{equation}\label{eq:pauc}
%\frac{1}{\sum_{i=1}^{M}m_{\ell i}\sum_{j=1}^Jn_{\ell j} }\sum_{i=1}^M\sum_{p=1}^{m_{\ell i}}\sum_{j=1}^{J} \sum_{q=1}^{n_{\ell j}}I(X_{\ell i p}> Y_{\ell j q}|Y_{\ell j q}\in (\hat S^{-1}_{\bar D,\ell}(u_2), \hat S^{-1}_{\bar D,\ell}(u_1)).
%\end{equation}
The pAUC statistic is useful in disease screening when a high FPR  would lead to a large number of falsely diagnosed subjects. It is  desirable to evaluate and compare the marker accuracy at the low FPRs rather than the entire range of FPRs.   When we are interested in the sensitivity of the $\ell$th marker at a particular threshold, say $c$, we can specify the probability measure to be    a point mass at $u_0=S_{\bar D,\ell} (c)$.  The estimator $\widehat{\Omega}_\ell$ then becomes
\begin{equation}\label{eq:sen}
\frac{1}{\sum_{i=1}^{M}m_{\ell i}} \sum_{i=1}^M\sum_{p=1}^{m_{\ell i}} I(X_{\ell i p}> Y_{[(1-u_0)\sum_{j=1}^Jn_{\ell j}]}) .
\end{equation} The estimator in (\ref{eq:sen}) is obtained by comparing  all X's with $ Y_{[(1-u_0)\sum_{j=1}^Jn_{\ell j}]}$.

%The derivation of the asymptotic normality of $\boldsymbol{\widehat{\Omega}}$ is given in the Appendix, along with the explicit expression of the covariance matrix $\boldsymbol\Sigma$ for $\boldsymbol{\widehat{\Omega}}$.
%In the Appendix, we show that $\boldsymbol{\widehat{\Omega}}$ is asymptotically normal, and its variance expression is given by
%\begin{equation}\label{eq:nomega}var(\sqrt{M}(\widehat{\boldsymbol{\Omega}} ))\equiv\boldsymbol\Sigma=\boldsymbol\Sigma_1+\boldsymbol\Sigma_2, \end{equation} where the $\{\ell_1,\ell_2\}$ element   in  $\boldsymbol\Sigma_1$ is given by  \begin{equation}\label{eq:covw}  \alpha_{\ell_1}\alpha_{\ell_2}    \eta^x_{\ell_1,\ell_2} \int_0^1\!\!\!\int_0^1   [ S_{D,\ell_1,\ell_2}\{S_{\bar D,\ell_1}^{-1}(s),S_{\bar D,\ell_2}^{-1}(t)
%\}-S_{D,\ell_1}  \{S_{\bar D,\ell_1}^{-1}(s)\}S_{D,\ell_2}
% \{S_{\bar D,\ell_2}^{-1}(t)\}]dW(s)dW(t), \end{equation} and the   $\{\ell_1,\ell_2\}$ element in $\boldsymbol\Sigma_2$ is \begin{equation} \label{eq:covv}     {\lambda\beta_{\ell_1}  {\beta_{\ell_2} }}
%  \eta^y_{\ell_1,\ell_2}\int_0^1\!\!\!\int_0^1
% r_{\ell_1}(s)r_{\ell_2}(t)[S_{\bar D,\ell_1, \ell_2}\{S_{\bar D,\ell_1}^{-1}(s),S_{\bar D,\ell_2}^{-1}(t)
%\}-st ]dW(s)dW(t). \end{equation}

In the following sections, we propose efficient nonparametric methods based on the  nonparametric estimator of $\boldsymbol{\Omega}$ to evaluate and compare multiple  markers in multi-reader multi-test ROC Data and longitudinal ROC data.

\subsection{Multi-reader multi-test ROC data}\label{sec:mrmt}
%We consider applying the $\Delta$-statistic to various types of correlated  marker data by assigning different functions $h$. Our nonparametric statistics do not require model assumptions, offering an advantage over existing methods for these types of data. In the following sections, we will discuss two common correlated biomarker data including multi-reader multi-test ROC data and longitudinal ROC data.
\textbf{One type of complex marker data arise frequently in medical imaging studies when  radiological images of a patient are evaluated by  several radiologists.   \cite{obuchowski:95}  consider  a mixed-effect ANOVA model while allowing for correlation among AUC estimators.  Their model requires a specific covariance structure  among the AUCs.  \cite{song:05}    propose   a pseudo-generalized estimating equation   method  and derive   large sample theory for the estimators. Their method remains valid under the working-independence assumption.}

In a   multi-reader multi-test ROC study, suppose  the  radiologist $r$, $r=1,...,R$,  rates images for $M$ diseased subjects and   $J$ non-diseased subjects from $L$ imaging devices.   A radiologist can give one or more ratings to   suspicious locations   in each subject, that is, $m_{\ell i}, n_{\ell j}\geq 1.$ We consider $L=2$. Denote $ \Omega_1, ..., \Omega_R $ as wAUCs from $R$ readers for modality 1, $\Omega_{R+1}, ..., \Omega_{2R}$ as wAUCs from $R$ readers for modality 2.  Common nonparametric approaches for comparing imaging modalities   take the difference $\Omega_{r}-  \Omega_{R+r}$ between two devices for   reader $r$, and then average these differences over all reader    \cite{lee:01}.  We can see that such methods are a special case of the linear combination of the weighted AUC statistics for reader-modality combinations.  Rather than the simple average of all $\Omega_{r}-  \Omega_{R+r}$'s, we propose to use the following weighted linear combination  to possibly achieve a higher power to compare markers
\begin{equation}\label{eq:hmm}
\Delta_m= (\sum_{r=1}^Rw_r)^{-1}\sum_{r=1}^R[w_r(\Omega_{r}-  \Omega_{R+r})],
\end{equation} with positive and bounded weights $\tilde{W}=(w_1, w_2, ...w_R)'.$   The parameter $\Delta_m$ can be empirically estimated by  \[\hat\Delta_m= (\sum_{r=1}^Rw_r)^{-1}\sum_{r=1}^R[w_r(\widehat{\Omega}_{r}-  \widehat{\Omega}_{R+r})] ,\] which compares two modalities with multiple readers.

Various choices of weights exist in the ROC literature.   $\tilde{W}$ may not depend on the data. For instance, if all readers are assumed to be homogeneous with regard to their accuracy of rating images,   an equal weight $w_r=1/R$ can be assigned to  reader $r$,  $r=1,...,R$. Then with $m_{\ell i}= n _{\ell j}=1$ and $W(u)=1$ at $0<u<1$, $\hat\Delta_m$ becomes the  AUC statistic in \cite{lee:01}.   When one has to estimate $\tilde{W}$  from the data, the consistency  of estimated weights $\hat{W}$ in probability is required for the derivation.    For instance, a set of optimal  weights   is introduced by \cite{wei:85} and further developed by \cite{jin:06}, who argues that when readers' experience vary greatly, using equal weights may yield a biased AUC estimate. Let the $R\times R$ covariance matrix of estimated AUC differences, $(\widehat{\Omega}_{1}-  \widehat{\Omega}_{R+1}, ..., \widehat{\Omega}_{R}-  \widehat{\Omega}_{2R})'$,  be ${\boldsymbol\Sigma}_A$, and its consistent estimator $\hat{{\boldsymbol\Sigma}}_A$.  They then choose   $\tilde{W}= \hat{{\boldsymbol\Sigma}_A^{-1}}\textbf{1}$  to obtain a consistent estimator for the AUC difference, where   $ \textbf{1}$ is a $R$-dimensional vector of one's.  \cite{wei:85} and \cite{jin:06} show that this set of weights are optimal since they maximize the local power to detect the AUC difference between imaging modalities. It is clear that by combining these weights with $m_{\ell i}= n _{\ell j}=1$ and $W(u)=1$ at $0<u<1$, $\hat\Delta_m$ becomes  \cite{jin:06}'s statistic. To properly calculate the weights for the proposed statistic, we need to obtain the covariance matrix $\boldsymbol\Sigma$ of $\boldsymbol{\widehat{\Omega}}=(\widehat{\Omega}_{1},...,\widehat{\Omega}_{2R})'$. Since in practice $\boldsymbol\Sigma$  is unknown, its consistent estimator $\widehat{\boldsymbol\Sigma}$ can be obtained using the explicit expression (\ref{eq:nomega}) derived     in the Appendix. Since $\boldsymbol\Sigma$ and ${\boldsymbol\Sigma}_A$ is related via \[{\boldsymbol\Sigma}_A=\boldsymbol\Sigma \boldsymbol A,\] where the $r$th column of the $2R\times R$ matrix $\boldsymbol A$ has 1's at $r$th and $(R+r)$th rows and 0 at other rows,    the estimated weights are given by \begin{equation}\label{eq:optiw}
\hat{W}= {\widehat{\boldsymbol\Sigma}} ^{-1}\boldsymbol A\textbf{1}.
\end{equation}

\subsection{Longitudinal  biomarker data}\label{sec:logi}

Another example of complex marker data comes from longitudinal studies when marker measurements are taken at several times during the studies. Most methodology for longitudinal ROC data   rely on appropriate assumptions on the distributions of marker measurements  \cite{Etzioni:99}. In  longitudinal ROC data, suppose $L$ markers are measured on $M$ diseased patients and $J$ non-diseased patients at times   $t_1, t_2, ..., t_K$.
Suppose each subject is repeatedly measured    for every marker at each time. Let $X_{\ell i p k}$ denote the test result of marker $\ell$  in the $p$th repetition  on the diseased subject $i$ at time $t_k$, where $\ell=1,...,L$, $p  = 1,...,m _{\ell ik}$,   $i=1,...M$, and $k=1,...,K$.  Let $Y_{\ell j q k}$ denote test result of $\ell$th marker on the $p$th repetition  in the non-diseased subject $j$ at time $t_k$, where $\ell=1,...,L$, $q = 1,...,n _{\ell jk}$,    $j=1,...J$, and $k=1,...,K$.    The nonparametric wAUC estimator for the $\ell$th marker is then given by $\widehat{\Omega}_\ell=
\int_0^1{ \hat {S}_{D,\ell}}\{{ \hat {S}_{\bar D,\ell}}^{-1}(u)\} dW(u),$ where   $\hat S_{D,\ell}$ and $\hat S_{\bar D,\ell}$    are defined   by
\begin{eqnarray}\label{eq:fcom}\hat S_{D,\ell} (x)=\frac{1}{\sum_{i=1}^{M}\sum_{k=1}^{K}m_{\ell i k}}\sum_{i=1}^M\sum_{k=1}^{K}\sum_{p=1}^{m_{\ell i k}}I(X_{\ell i pk}> x ), \nonumber\\
 \textrm{and}\quad \hat S_{\bar D,\ell} (x)=\frac{1}{\sum_{j=1}^{J}\sum_{k=1}^{K}n_{\ell jk}}\sum_{j=1}^J\sum_{k=1}^{K}\sum_{q=1}^{n_{\ell j k}}I(Y_{\ell j qk}> x).
 \end{eqnarray}
By defining $W(u)$ accordingly in the wAUC estimator, we obtain
the nonparametric AUC  estimator for the $\ell$th marker:
\begin{equation} \frac{1}{\sum_{i=1}^{M}\sum_{k=1}^{K}m_{\ell i k}\sum_{j=1}^{J}\sum_{k=1}^{K}n_{\ell jk}}\sum_{i=1}^M\sum_{k_1=1}^{K}\sum_{p=1}^{m_{\ell i k}}\sum_{j=1}^J\sum_{k_2=1}^{K}\sum_{q=1}^{n_{\ell j k}}I(X_{\ell i p k_1}> Y_{\ell j q k_2}),\nonumber
\end{equation}
the partial AUC estimator:
\begin{equation}
\frac{\sum_{i=1}^M\sum_{k_1=1}^{K}\sum_{p=1}^{m_{\ell i k}}\sum_{j=1}^J\sum_{k_2=1}^{K}\sum_{q=1}^{n_{\ell j k}}I(X_{\ell i p k_1}> Y_{\ell j q k_2}|Y_{\ell j q k_2}\in (\hat S^{-1}_{\bar D,\ell}(u_2), \hat S^{-1}_{\bar D,\ell}(u_1))}{\sum_{i=1}^{M}\sum_{k=1}^{K}m_{\ell i k}\sum_{j=1}^{J}\sum_{k=1}^{K}n_{\ell jk}},\nonumber
\end{equation}
and the sensitivity estimator at the FPR of $u_0$,
\begin{equation}
\frac{1}{\sum_{i=1}^{M}\sum_{k=1}^{K}m_{\ell i k}} \sum_{i=1}^M\sum_{k =1}^{K}\sum_{p=1}^{m_{\ell i k}} I(X_{\ell i p k}> Y_{[(1-u_0)\sum_{j=1}^{J}\sum_{k=1}^{K}n_{\ell jk}]}) .\nonumber
\end{equation}

We define $h$ to be a  real-valued   function of $\boldsymbol{\widehat{\Omega}}$. Here the function $h$ is defined on $\mathbb{R}^L$, and has continuous partial derivatives of order 2.    Let the   ROC summary measure   be $ \Delta_h =  h(\boldsymbol{ \Omega})$. Its   empirical estimator is given by
\begin{equation}\label{eq:delta}
\hat\Delta_h \equiv h(\boldsymbol{\widehat{\Omega}})=h\left(\int_0^1{ \hat {S}_{D,1}}\{{ \hat {S}_{\bar D,1}}^{-1}(u)\} dW(u),  ..., \int_0^1{ \hat {S}_{D,L}}\{{ \hat {S}_{\bar D,L}}^{-1}(u)\} dW(u)\right).
\end{equation}

 The statistic above can be used to compare two longitudinal markers   when $h$ is a  linear contrast.
 $\hat\Delta_h$ also includes a broad range of ROC statistics. It is the weighted AUC statistic in   \cite{wgjj:89} and later in \cite{li:10} for evaluating and comparing markers. When $W(u)=1$ at $0<u<1$ and $h$ is a linear function,  $\hat\Delta_h$ is the generalized AUC statistic  in \cite{lee:01}. When  $W(u)=1$ at $0<u<1$, $\hat\Delta_h$ is the AUC statistic in \cite{gang:08}, assuming no correlation between $X$ and $Y$, which    allows for multiple observations per patient from each marker. When $W(u)=(u-a)/(b-a)$ for $0<a<u<b<1$ and
  $h(\Omega_1,\Omega_2 )=\Omega_1-\Omega_2$, $\hat\Delta_h$ is the pAUC statistic in \cite{zhang:02} for comparing two markers.

  \textbf{{{When there are two longitudinal markers in the study,  the optimal combination for comparing the two markers can be obtained using the  similar steps in the aforementioned  multi-reader multi-test studies.  Suppose $L=2$. Let ${\Omega}_{\ell,k}$ be the wAUC of marker $\ell,\ell=1,2,$ at time $t_k$    and $\widehat{\Omega}_{\ell,k}$ be its nonparametric estimator  given by $\widehat{\Omega}_{\ell,k}=
\int_0^1{ \hat {S}_{D,\ell,k}}\{{ \hat {S}_{\bar D,\ell,k}}^{-1}(u)\} dW(u),$ where   $\hat S_{D,\ell,k}$ and $\hat S_{\bar D,\ell,k}$    are defined   by
\begin{eqnarray}\hat S_{D,\ell,k} (x)=\frac{1}{\sum_{i=1}^{M}m_{\ell i k}}\sum_{i=1}^M\sum_{p=1}^{m_{\ell i k}}I(X_{\ell i pk}> x ),
 \textrm{and}\quad \hat S_{\bar D,\ell,k} (x)=\frac{1}{\sum_{j=1}^{J}n_{\ell jk}}\sum_{j=1}^J\sum_{q=1}^{n_{\ell j k}}I(Y_{\ell j qk}> x).
 \end{eqnarray}
Note that the estimation of  ${\Omega}_{\ell,k}$ is based on every individual time point. One can take difference of the wAUCs of two markers, and  simply average these differences over all time points. We may also  use the following weighted linear combination  to possibly achieve a higher power to compare markers
\begin{equation}
\Delta_{\ell}= (\sum_{k=1}^Kw_k)^{-1}\sum_{k=1}^K[w_k(\Omega_{1,k}-  \Omega_{2,k})],
\end{equation} with positive and bounded weights $\tilde{W}=(w_1, w_2, ...w_K)'.$   The parameter $\Delta_{\ell}$ can be empirically estimated by  \[\hat\Delta_{\ell}= (\sum_{k=1}^Kw_k)^{-1}\sum_{k=1}^K[w_k(\widehat{\Omega}_{1,k}-  \widehat{\Omega}_{2,k})] .\]
 Similarly as in  the previous section,   the $2K\times 2K$ covariance matrix $\boldsymbol\Sigma$ of $\boldsymbol{\widehat{\Omega}}=(\widehat{\Omega}_{1,k},...,\widehat{\Omega}_{2K})'$ can be estimated can be obtained using the explicit expression in (\ref{eq:nomega}). Thus the estimated weights are given by the same expression as  (\ref{eq:optiw}).
}}}

 \section{Asymptotic variance expressions of  the proposed statistics}\label{sec:deltas}
In this section we derive the asymptotic variances for the proposed statistics in the multi-reader multi-test data and the longitudinal data.  We first show the explicit variance expressions for $\hat\Delta_m$, and then show the variance expression for the more general statistic $\hat\Delta_h$ in (\ref{eq:delta}) for the longitudinal data.

   The numbers of abnormal locations within a diseased subject may differ, and so are the numbers of normal locations within a non-diseased subject.  Denote $\tilde{m}_\ell=\sum_{i=1}^{M} {m}_{\ell i}$, and $\tilde{n}_\ell=\sum_{j=1}^{J} {n}_{\ell j}$.
   Assume that  $S_{D,\ell}$ and $S_{\bar D,\ell}$ have continuous and positive  derivatives, $S'_{D,\ell}$, and $S'_{\bar D,\ell}$.   In Appendix  we  show that the proposed statistic,  $\hat\Delta_m$,   for   the multi-reader multi-test ROC data  is  asymptotically normal
when sample sizes are large. The variance of $\hat\Delta_m$ has the following expression   when sample sizes get large:
        \begin{equation}\label{eq:varm}
        var(\hat\Delta_m ) =\tilde{v}_X+   \tilde{v}_Y   ,
        \end{equation}
with
    \begin{align}
   \tilde{v}_X \!\!=\!\!   \frac{1}{M\tilde{m}_{\ell_1}\tilde{m}_{\ell_2}(\sum_{r=1}^Rw_r)^{ 2}}    \sum_{1\leq\ell_1,\ell_2\leq 2R}        \sum_{i=1}^M     &\tilde{m}_{\ell_1  i} \tilde{m}_{\ell_2  i}   (-1)^{I(\ell_1,\ell_2)+1}
  \Big( \int \!\!\!\int    \big [ S_{D,\ell_1,\ell_2}\{S_{\bar D,\ell_1}^{-1}(s),S_{\bar D,\ell_2}^{-1}(t)
\}\nonumber\\
&  -S_{D,\ell_1}  \{S_{\bar D,\ell_1}^{-1}(s)\}S_{D,\ell_2}
 \{S_{\bar D,\ell_2}^{-1}(t)\}\big]dW(s)dW(t)\Big),\nonumber
 \end{align}
 and
    \begin{align}
 \tilde{v}_Y  \!\!=\!\!    \frac{1}{M\tilde{n}_{\ell_1}\tilde{n}_{\ell_2}(\sum_{r=1}^Rw_r)^{ 2}}     \sum_{1\leq\ell_1,\ell_2\leq 2R} \sum_{j=1}^J    &\tilde{n}_{\ell_1  j} \tilde{n}_{\ell_2  j}   (-1)^{I(\ell_1,\ell_2)+1}  \Big( \int \!\!\!\int
 r_{\ell_1}(s)r_{\ell_2}(t)\nonumber\\ \times\big[S_{\bar D,\ell_1, \ell_2}&\{S_{\bar D,\ell_1}^{-1}(s),S_{\bar D,\ell_2}^{-1}(t)
\}-st \big]dW(s)dW(t)\Big),\nonumber
\end{align} where $I(\ell_1,\ell_2)=1, $ if $ |\ell_2-\ell_1 |< R$, and 0, otherwise, and
\begin{displaymath}
r_\ell (u)=  S'_{D,\ell} \{S_{\bar D,\ell}
^{-1}( u)\}/S'_{\bar D,\ell} \{S_{\bar D,\ell} ^{-1}( u)\} , \quad \mbox{for}\quad \ell=1,...,L.
\end{displaymath}
 The marginal and joint survivor functions can also be empirically estimated.

   Denote $m_\ell=\sum_{i=1}^{M}\sum_{k=1}^{K}m_{\ell i k}$, and $n_\ell=\sum_{j=1}^{J}\sum_{k=1}^{K}n_{\ell j k}$.    we  show in Appendix that the proposed statistic,  $\hat\Delta_h$ in (\ref{eq:delta}) for the longitudinal data is also asymptotically normal, and the variance of $\hat\Delta_h$ takes on the  following form
when sample sizes are large,
  \begin{equation}\label{eq:varm}
        var(\hat\Delta_h  ) = {v}_X+    {v}_Y   ,
        \end{equation}
         where
  \begin{align}
   v_X \!\!=\!\!      \frac{1}{M {m}_{\ell_1} {m}_{\ell_2} }           \sum_{i=1}^M     {m}_{\ell_1  i}  {m}_{\ell_2  i}& \frac{\partial{h}}{\partial{\Omega_{\ell_1}}} \frac{\partial{h}}{\partial{\Omega_{\ell_2}}}
  \Big( \int \!\!\!\int    [ S_{D,\ell_1,\ell_2}\{S_{\bar D,\ell_1}^{-1}(s), S_{\bar D,\ell_2}^{-1}(t)
\}\nonumber\\
&  -S_{D,\ell_1}  \{S_{\bar D,\ell_1}^{-1}(s)\}S_{D,\ell_2}
 \{S_{\bar D,\ell_2}^{-1}(t)\}]dW(s)dW(t)\Big),\nonumber
 \end{align}
 and
    \begin{align}
 v_Y \!\!=\!\!       \frac{1}{M {n}_{\ell_1} {n}_{\ell_2} }       \sum_{j=1}^J     {n}_{\ell_1  j}  {n}_{\ell_2  j}  \frac{\partial{h}}{\partial{\Omega_{\ell_1}}}\frac{\partial{h}}{\partial{\Omega_{\ell_2}}}   \Big( \int \!\!\!\int
 r_{\ell_1}(s)r_{\ell_2}(t)[S_{\bar D,\ell_1, \ell_2}\{S_{\bar D,\ell_1}^{-1}(s),S_{\bar D,\ell_2}^{-1}(t)
\}-st ]dW(s)dW(t)\Big),\nonumber
\end{align}
where
\begin{displaymath}
r_\ell (u)=  S'_{D,\ell} \{S_{\bar D,\ell}
^{-1}( u)\}/S'_{\bar D,\ell} \{S_{\bar D,\ell} ^{-1}( u)\} , \quad \mbox{for}\quad \ell=1,...,L.
\end{displaymath}
 {The empirical or other type of smoothed estimators for the marginal and joint survivor functions $  {S}_{D,\ell} $, $ {S}_{\bar D,\ell}$, $S_{D,\ell_1, \ell_2}(x_1,x_2)$, and $ S_{\bar D,\ell_1, \ell_2}(y_1,y_2)$ can be used to estimate $ {v}_X$ and
$   {v}_Y$. In the simulations and the example, we used the empirical estimators. That is, we estimate $  {S}_{D,\ell} $ and  $ {S}_{\bar D,\ell}$ using the expressions in (\ref{eq:fcom}). And we estimate $S_{D,\ell_1, \ell_2}(x_1,x_2)$, and $ S_{\bar D,\ell_1, \ell_2}(y_1,y_2)$ as   follows:
\begin{align}
 \hat S_{D,\ell_1, \ell_2}(x_1,x_2)&=\frac{1}{\sum_{i=1}^{M}m^2_{\ell i}}\sum_{i=1}^M\sum_{p_1=1}^{m_{\ell_1 i}}\sum_{p_2=1}^{m_{\ell_2 i}}\sum_{k_1=1}^{K}\sum_{k_2=1}^{K}I(X_{\ell_1 i p_1 k_1}> x_1,   X_{\ell_2 i p_2 k_2}> x_2), \nonumber\\
 \hat  S_{\bar D,\ell_1, \ell_2}(y_1,y_2)&= \frac{1}{\sum_{j=1}^{J}n^2_{\ell j}}\sum_{j=1}^J\sum_{q_1=1}^{n_{\ell_1 i}}\sum_{q_2=1}^{n_{\ell_2 j}}\sum_{k_1=1}^{K}\sum_{k_2=1}^{K}I(Y_{\ell_1 j q_1 k_1}> y_1,  Y_{\ell_2 j q_2 k_2}> y_2) .\nonumber
\end{align}}
Thus, when $\Omega$'s are AUCs, $v_X$ is given by
\begin{align}
   v_X \!\!=\!\!      \frac{1}{M {m}_{\ell_1} {m}_{\ell_2} }       \sum_{1\leq\ell_1,\ell_2\leq 2R}     \sum_{i=1}^M     {m}_{\ell_1  i}  {m}_{\ell_2  i}& \frac{\partial{h}}{\partial{\Omega_{\ell_1}}} \frac{\partial{h}}{\partial{\Omega_{\ell_2}}}
  \Big(      E[I(X_{\ell_1 i p_1 k_1}>Y_{\ell_1 j p_1 k_1})I(X_{\ell_2 i p_1 k_1}>Y_{\ell_2 j p_1 k_1})]
 \nonumber\\
&  -E[I(X_{\ell_1 i p_1 k_1}>Y_{\ell_1 j p_1 k_1})]E[I(X_{\ell_2 i p_1 k_1}>Y_{\ell_2 j p_1 k_1})]\Big),\nonumber
 \end{align}
 and $v_Y$ is given by
    \begin{align}
 v_Y \!\!=\!\!       \frac{1}{M {n}_{\ell_1} {n}_{\ell_2} }    \sum_{1\leq\ell_1,\ell_2\leq 2R}    \sum_{j=1}^J     {n}_{\ell_1  j}  {n}_{\ell_2  j}  & \frac{\partial{h}}{\partial{\Omega_{\ell_1}}}\frac{\partial{h}}{\partial{\Omega_{\ell_2}}}   \Big(  E[I(X_{\ell_1 i p_1 k_1}>Y_{\ell_1 j p_1 k_1})I(X_{\ell_2 i p_1 k_1}>Y_{\ell_1 j p_1 k_1})]
 \nonumber\\
&  -E[I(X_{\ell_1 i p_1 k_1}>Y_{\ell_1 j p_1 k_1})]E[I(X_{\ell_2 i p_1 k_1}>Y_{\ell_2 j p_1 k_1})]\Big),\nonumber
\end{align}

\section{Simulation studies}\label{sec:sim}
 We report  simulation studies to evaluate the finite sample property of the proposed statistics. We simulated  both  multi-reader multi-test ROC data and longitudinal data. In multi-reader multi-test data, we considered the finite sample performance of the variance expression. More importantly, we compared the simulated powers of the equal weight and the optimal weight  introduced in Section \ref{sec:mrmt}. We expect that the optimal weight results in better power than the equal weight. In longitudinal data we considered the general setting where each subject is diagnosed repeatedly at each time point and the number of  repeated measures   varies from subject to subject.

 \subsection{Multi-reader multi-test data}
 In the first simulation study we investigated the  finite sample accuracy of the variance expression   for multireader multitest data.  We   let $m_{\ell i}= n _{\ell j}=1$, $R=3$, and $L=2$. We simulated 1000 datasets   under multivariate  normal and lognormal  distributions:
 \begin{enumerate}
 \item $X\sim N(\boldsymbol{\mu}_X,\boldsymbol\Sigma_X)$ and $Y\sim N(\boldsymbol{\mu}_Y,\boldsymbol\Sigma_Y)$, where $\boldsymbol{\mu}_X= (1,...,1), \boldsymbol{\mu}_Y=(0,...,0)$ and  $\boldsymbol\Sigma_X=\boldsymbol\Sigma_Y$ is the variance-covariance matrix with diagonal  elements $(1,1.5,2,1,1.5,2)$ and correlation coefficient, $\rho$;
 \item $X\sim LogNormal(\boldsymbol{\mu}_X,\boldsymbol\Sigma_X)$ and $Y\sim LogNormal(\boldsymbol{\mu}_Y,\boldsymbol\Sigma_Y)$.
 \end{enumerate}
  From simulated data we used    the proposed statistic in Section \ref{sec:mrmt}, $\hat{\Delta}_m=\sum_{r=1}^3(\hat\Omega_{r}-  \hat\Omega_{R+r})/R$ to estimate  the AUC by defining the weight function $W(u)=1,$ for $0<u<1$), and the pAUC  by defining $W(u)=1,$ for $0<u<0.6$; 0 otherwise. A 95\% confidence interval for $\hat{\Delta}_m$ was obtained using the variance expression derived in (\ref{eq:varm}). Table \ref{tb:mmd} shows   biases, square root of mean squared errors (RMSE), and simulated coverage of confidence intervals.  It is clear from the table that coverage levels are   close to the nominal level, and   biases for comparing AUCs or pAUCs are close to zero.   This shows good performance of our estimator and associated asymptotic results.

 \textbf{In the second simulation study we compared the performance of the proposed method with the parametric method by \cite{molodianovitch:06} and the semiparametric logistic regression method by \cite{Copas:02} with regard to estimating the AUC. We used the same setting as the first simulation study except changing $\boldsymbol{\mu}_X$ to $(1, 1, 1, 1.5,2,2.5)$. The biases and RMSEs from the three methods are shown in Table \ref{tb:mmdc}.   {{The results indicate that the proposed method and the semiparametric method perform much better than the parametric method when the distribution assumptions are violated. They also indicate that the semiparametric method performs as well as the proposed method. This is not surprising as can be seen from the   description of the semiparametric method in Section 2 of \cite{Copas:02}. The logistic regression fits the regression parameters based on the following equation: \[logit(D=1)=\beta_0+\beta_1 Z,\] where $D$ is the disease status (with 1 being the diseased, and 0 being the non-diseased), $\beta_0$ and $\beta_1$ are regression parameters, and $Z$ is the test result.  After the regression parameter estimators, $\hat{\beta}_0$ and $\hat{\beta}_1$, are obtained,   the empirical ROC curve is estimated based on the new score, $\tilde{Z}= \hat{\beta}_0+\hat{\beta}_1 Z$. Since the ROC curve is invariant to monotonic transformation, the empirical ROC curve based on the new score remains the same as the empirical ROC curve from the original test results. }}}

 In the third simulation study we compared the simulated powers using the optimal weight versus  the equal weight.   We again  let $m_{\ell i}= n _{\ell j}=1$, $R=3$, and $L=2$. We simulated 1000 datasets   under multivariate  normal distributions:
 $X\sim N(\boldsymbol{\mu}_X,\boldsymbol\Sigma_X)$ and $Y\sim N(\boldsymbol{\mu}_Y,\boldsymbol\Sigma_Y)$, where $\boldsymbol{\mu}_X= (2,1,...,1), \boldsymbol{\mu}_Y=(0,...,0)$ and  $\boldsymbol\Sigma_X=\boldsymbol\Sigma_Y$ is the variance-covariance matrix with diagonal  elements $(1,1.5,2,2,3,2)$ and correlation coefficient, $\rho$.
 We selected $m=n$ in (50,100), and $\rho$ in $(-0.1,0.2,0.5)$. For each simulated data, we estimated the weighted  differences in (\ref{sec:mrmt}):
 \[h( \boldsymbol{\Omega})= (\sum_{r=1}^3w_r)^{-1}\sum_{r=1}^3[w_r(\Omega_{r}-  \Omega_{3+r})],\]   with both equal weights ($w_r=1/3$) and the optimal weights given in (\ref{eq:optiw}). The AUC was estimated by defining the weight function $W(u)=1,$ for $0<u<1$), and the pAUC  was estimated by defining $W(u)=1,$ for $0<u<0.6$; 0 otherwise.  The simulated power was then calculated as  the number of rejections out of 1000 simulated datasets. Table \ref{tb:optip} shows   the simulated powers for the comparison of AUCs and pAUCs.  It is clear   that the optimal weights always  result in much larger powers than the equal weights.

  \subsection{Longitudinal biomarker data}

In this simulation study we generated multivariate log-normal correlated biomarker data. We generated data by taking exponential of multivariate normal data  $\boldsymbol{X}_i\sim N(\boldsymbol{\mu}_{X,i},\boldsymbol\Sigma_{X,i}) $ and $\boldsymbol{Y}_j\sim N(\boldsymbol{0} ,\boldsymbol\Sigma_{Y,j})$, where $\boldsymbol{\mu}_{X,i}= (2,...,2,1,...,1) $, and $ \boldsymbol\Sigma_{X,i}$ and $\boldsymbol\Sigma_{Y,j}$ are variance-covariance matrices. We let $L=2$, $K=3$, $M=J=(50, 200)$. To allow various cluster sizes, we let $m _{\ell ik}=2$ for the first half of diseased subjects,    and $m _{\ell ik}=4$ for the other half. For non-diseased subjects, let $n _{\ell jk}=5$ for the first half, and $n_{\ell jk}=3$ for the other half. We chose $\boldsymbol\Sigma_{X,i}=(1-\rho)\boldsymbol{M}_i+\rho\boldsymbol{1}_i\boldsymbol{1}_i'$, where  $\boldsymbol{M}_i$ is the $LKm_{\ell ik}\times LKm_{\ell ik}$ identity matrix and $\boldsymbol{1}_i$ is the $LKm_{\ell ik}\times 1$   matrix  with all elements 1. Similar setting was applied to define $\boldsymbol\Sigma_{Y,j}$.  Here $\rho$ gives  within-subject correlation. We let $\rho=0.4$ for the diseased  and $\rho=0.3$ for the non-diseased. We simulated 1000 datasets for each sample size, and obtained   the estimate of AUC difference between two biomarkers, $\hat{\Delta}_l$, and its variance. Table \ref{tb:lbd} shows   biases, square root of mean squared errors (RMSE), and simulated coverage of confidence intervals.     This again shows good performance of our estimator   for correlated biomarker data.

\section{An example in the diagnosis of endometriosis}\label{sec:example1}
The proposed nonparametric ROC summary statistics are applied in this section to data from a study on endometriosis
diagnosis. Endometriosis is a gynecological medical condition in which endometrial-like cells appear and flourish in
areas outside the uterine cavity and is typically seen in women at their reproductive ages. It has been estimated that
endometriosis occurs in roughly 5\%--10\% of women. Despite its relatively high prevalence, substantive and
methodological challenges exist, including diagnostic proficiency. The Physician Reliability Study, an add-on to the Endometriosis: Natural History, Diagnosis and Outcome (ENDO) Study
\cite{bucklouis:11}, addressed this issue by investigating whether sequentially added clinical information of a
subject can aid in more accurately diagnosing the disease of endometriosis. Detailed study designs of ENDO and PRS can
be found in the aforementioned references. For demonstration purpose in this paper, we used review results of 4
physicians (reviewers) in PRS on 150 participants. All 150 participants had recorded operative digital images of their
pelvic organs and descriptive drawings and notes, both from surgeons who conducted the laparoscopies on these women in
ENDO study. The reviewers conducted their reviewing and diagnosis under two modalities. Modality one corresponds to the
setting where the reviewers are presented with participants' digital video/images while modality two corresponds to the
setting where both digital video/images and surgeon's reports (drawings and notes) are presented. For each participant
under each modality, the reviewer answered a series questions on what they observe from the clinical information. These
answered were later used to derive the rASRM scores \cite{asrm:97} which we used as the diagnostic outcomes in this
paper. The visualized diagnosis from the original ENDO study of these participants were used as the gold standard.

For the first modality, the estimated AUCs are $(0.71 , 0.75 , 0.63, 0.76 )$ for the four reviewers; the corresponding
numbers are $(0.83 , 0.85 , 0.75 , 0.87 )$ for the second modality.  With equal weights $w_r=1/4, r=1,...,4$, the
$\Delta$-statistic is $\hat\Delta_m= -0.1145,$ and its variance estimate is 0.0007475.  We used (\ref{eq:optiw}) to obtain the
optimal weights $(w_1,w_2,w_3,w_4)$=(298.08, 401.16 , 176.88, 560.48).  Using these weights, the
$\Delta$-statistic is given by $\hat\Delta_m= -0.1115,$ and its variance estimate is 0.0006961. This indicates that the
$\Delta$-statistic is more precisely estimated by using the optimal weights.  \textbf{ The   two-sided $p$-value using the optimal weights is $2.36\times 10^{-5}$, which is slightly smaller than the $p$-value $2.82\times 10^{-5}$ using
equal weights. } The two-sided $p$-values based on both sets of weights are both close to zero, which indicates that these physicians are able to give more precise
diagnosis on endometriosis by reviewing both digital images and surgeons' descriptive reports.

\section{Discussion}\label{sec:dis}
 The proposed  methods in the paper  are  nonparametric  and   can be applied to evaluate and  compare diagnostic markers  in the multireader multitest data and the longitudinal data. As illustrated in the simulation studies and the example, the proposed weighted method in the multireader multitest data tends  to have a larger power than the existing methods. We also conducted simulation studies to investigate  the finite sample performance of the proposed method in the longitudinal data setting. More complex correlated data in which both normal and abnormal locations may occur in the same subject  have been considered in \cite{brunner:07} and \cite{brunner:09}. How to extend the proposed statistics to such a data setting is a future research topic.

\textbf{{{As pointed out by a reviewer,  the proposed method is based on the empirical distribution estimators, and may not allow
  more complicated  dependencies of observations  in longitudinal data. For
example, in the case of  autoregressive
dependencies, empirical estimators could not converge to target
probabilities, especially when autoregression coefficients are greater than one.  More research is merited to extend the proposed method in this direction. }}}
 %Section \ref{sec:notation} discusses the   connection between the proposed AUC statistic and the Wilcoxon statistic in \cite{rosner:03}.  The variance expression in (\ref{eq:varm}) provides the asymptotic variance for the Wilcoxon statistic under both the balanced design and the unbalanced design.  For   marker $\ell$, we have the asymptotic variance of $A_\ell$:
% \begin{align}\label{eq:varw}
% var(A_\ell)=var(A_c)=&\tilde{m}_\ell \tilde{n}_\ell^2 \{P( X_{\ell i} >Y_{\ell j} ,X_{\ell i}>Y_{\ell j'})
% -P(X_\ell>Y_\ell)^2\}\nonumber\\
% &+ \tilde{m}_\ell^2\tilde{n}_\ell\{P( X_{\ell i} >Y_{\ell j} ,X_{\ell i'}>Y_{\ell j})-P(X_\ell>Y_\ell)^2\}.
% \end{align}
% The above variance  becomes $  \tilde{m}_\ell \tilde{n}_\ell(\tilde m_\ell +\tilde n_\ell)/12$  under the null, which is asymptotically equivalent to the result    in \cite{rosner:03} when no ties are present in the data. Moreover,  (\ref{eq:varw}) gives the variance expression under some alternative. Combining  this  variance and estimated $P(X_\ell>Y_\ell)$, we  get the power and sample sizes of the Wilcoxon test for correlated data.

\section*{Acknowledgement}
The authors would like to thank   an associate editor and two referees for their constructive
comments and suggestions. The project described  here was supported in part by Award Number R15CA150698 from the National Cancer Institute under the American Recovery and Reinvestment Act of 2009 and by Award Number H98230-11-1-0196 from the National Security Agency. The work was also supported in part with funding from the American Chemistry Council and the Intramural Research Program of the {\it Eunice Kennedy Shriver} National Institute of Child Health and Human Development.  The content is solely the responsibility of the authors and does not necessarily represent the official views of the National Cancer Institute or the National Institutes of Health.

\bibliographystyle{wileyj}
\bibliography{cite_genauc}

\section*{Appendix:  Derivation of variance expression  of $\Delta_h$ }
\renewcommand \theequation{A.\arabic{equation}}
\setcounter{equation}{0}
 Assume that  $S_{D,\ell}$ and $S_{\bar D,\ell}$ have continuous and positive  derivatives, $S'_{D,\ell}$, and $S'_{\bar D,\ell}$. Suppose that
$ M/ m_\ell \rightarrow \alpha_\ell$,  $M/ n_\ell \rightarrow \beta_\ell $,   $M/J\rightarrow \lambda$,
$ \sum_{i=1}^M    m_{\ell_1  i} m_{\ell_2  i} /M^2 \rightarrow\eta^X_{\ell_1,\ell_2} , $
and
$\sum_{j=1}^J     n_{\ell_1  j} n_{\ell_2  j}/M^2  \rightarrow   \eta^Y_{\ell_1,\ell_2} ,$
as $M,J\rightarrow \infty. $ Assume that $\alpha_\ell$,   $  \beta_\ell $, $ \eta^X_{\ell_1,\ell_2}$ and $  \eta^Y_{\ell_1,\ell_2}$ are finite numbers. In addition, assume that  the function
  $h$ has continuous partial derivatives of order 2 at each point of an open set  $ (\boldsymbol{\Omega}-\boldsymbol{\epsilon}, \boldsymbol{\Omega}+\boldsymbol\epsilon)$, for $\boldsymbol\epsilon>0$. Let    \begin{displaymath}
r_\ell (u)=  S'_{D, \ell} \{S_{\bar D, \ell}
^{-1}( u)\}/S'_{\bar D, \ell} \{S_{\bar D, \ell} ^{-1}( u)\} , \quad \mbox{for}\quad \ell=1,...,L, \end{displaymath} where $S'_{D, \ell}$ and $S'_{\bar D, \ell}$ are the first derivatives of $S_{D, \ell}$ and $S_{\bar D, \ell}$, respectively.

%The proof of expression (\ref{ex:vargsim}) is provided in the Appendix.
The asymptotic normality  of $\boldsymbol{\widehat{\Omega}}$ is derived using results from \cite{gang:08}, which gives that
for  markers $1, ...,L$,
\[\sqrt{M}\left( {\begin{array}{*{20}c} {\widehat{ROC}_{ 1}(u)-{ROC}_{ 1}(u)}  \\ {\widehat{ROC}_{ 2}(u)-{ROC}_{ 2}(u)}  \\ \vdots  \\ {\widehat{ROC}_{L}(u)-{ROC}_{L}(u)}  \\ \end{array} } \right)   \xrightarrow{d} \left( {\begin{array}{*{20}c} {\sqrt{\alpha_1}\mathbb{U}_{1, 1}[  S_{D, 1} \{   S_{\bar D, _1} ^{-1}(u)\}] -\sqrt {\beta_1}    r_1(u) \mathbb{U}_{2,  1}(u)  }  \\ {\sqrt{\alpha_2}\mathbb{U}_{1, 2}[S_{D,  2} \{ S_{\bar D, 2} ^{-1}(u)\}] -\sqrt {\beta_2}    r_2(u) \mathbb{U}_{2,  2}(u)  }  \\ \vdots \\ {\sqrt{\alpha_L}\mathbb{U}_{1, L}[S_{D,  L} \{ S_{\bar D, L} ^{-1}(u)\}] -\sqrt {\beta_L}    r_L(u) \mathbb{U}_{2,  L}(u)  }  \\
 \end{array} } \right)\]
 where  $\mathbb{U}_{1,\ell}$ and $\mathbb{U}_{2,\ell}$ are   limiting Gaussian
 processes. Therefore, after some calculation, it follows that
\begin{equation}\label{eq:nomega}\sqrt{M}(\widehat{\boldsymbol{\Omega}}-\boldsymbol{\Omega})\xrightarrow{d} {\boldsymbol N}_{L}(\boldsymbol 0, \boldsymbol\Sigma=\boldsymbol\Sigma_1+\boldsymbol\Sigma_2), \end{equation} where the $\{\ell_1,\ell_2\}$ element   in  $\boldsymbol\Sigma_1$ is given by  \begin{equation}\label{eq:covw}  \alpha_{\ell_1}\alpha_{\ell_2}    \eta^x_{\ell_1,\ell_2} \int_0^1\!\!\!\int_0^1   [ S_{D,\ell_1,\ell_2}\{S_{\bar D,\ell_1}^{-1}(s),S_{\bar D,\ell_2}^{-1}(t)
\}-S_{D,\ell_1}  \{S_{\bar D,\ell_1}^{-1}(s)\}S_{D,\ell_2}
 \{S_{\bar D,\ell_2}^{-1}(t)\}]dW(s)dW(t), \end{equation} and the   $\{\ell_1,\ell_2\}$ element in $\boldsymbol\Sigma_2$ is \begin{equation} \label{eq:covv}     {\lambda\beta_{\ell_1}  {\beta_{\ell_2} }}
  \eta^y_{\ell_1,\ell_2}\int_0^1\!\!\!\int_0^1
 r_{\ell_1}(s)r_{\ell_2}(t)[S_{\bar D,\ell_1, \ell_2}\{S_{\bar D,\ell_1}^{-1}(s),S_{\bar D,\ell_2}^{-1}(t)
\}-st ]dW(s)dW(t). \end{equation}
The Taylor expansion of $\hat\Delta$ at $\boldsymbol\Omega$ gives
\begin{equation}\label{eq:varde1}
%\hat\Delta_h-\Delta_h\approx \frac{\partial h(\boldsymbol\Omega)}{\partial \Omega_1}(\hat \Omega_1-\Omega_1)+\frac{\partial h(\boldsymbol\Omega)}{\partial \Omega_2}(\hat \Omega_2-\Omega_2)+...+\frac{\partial h(\boldsymbol\Omega)}{\partial \Omega_1}(\hat \Omega_L-\Omega_L).\nonumber
\hat\Delta_h-\Delta_h\xrightarrow{d} (\widehat{\boldsymbol{\Omega}}-\boldsymbol{\Omega})'\nabla h(\boldsymbol{\Omega}),
\end{equation} where $\nabla h(\boldsymbol{\Omega})$ is the gradient of $h$ evaluated at $\boldsymbol{\Omega}$.
Since  the asymptotic variance of the  right hand side in (\ref{eq:varde1}) is given by
\begin{equation}\label{eq:vdelta}
  \nabla h(\boldsymbol{\Omega})'var(\widehat{\boldsymbol{\Omega}}-\boldsymbol{\Omega}) \nabla h(\boldsymbol{\Omega}).\nonumber
\end{equation}
 It follows that
\begin{equation}\label{eq:Avar}
var(\hat\Delta_h-\Delta_h)\xrightarrow{p}\sum_{\ell_1,\ell_2}  \frac{\partial{h}^2}{\partial{\Omega_{\ell_1}} \partial{\Omega_{\ell_2}}}
  cov(\hat \Omega_{\ell_1}-\Omega_{\ell_2},\hat \Omega_{\ell_1}-\Omega_{\ell_2}).\end{equation}
Using the covariance structures   in (\ref{eq:covw}) and (\ref{eq:covv}) in (\ref{eq:Avar}),  we can then obtain the asymptotic normality of $\hat\Delta_h$ by combining (\ref{eq:nomega}) with the Cramer-Wold device  \cite{serfling:80}.

\begin{table}[!htbp]
  \centering
  \caption{Bias, RMSE and  coverage for simulated multi-reader multi-test data}
    \begin{tabular}{cccccccccccccccccccccc}
\hline
          &       & &        \multicolumn{ 3}{c}{AUC} &       & \multicolumn{ 3}{c}{pAUC} \\
 \cline{4-6} \cline{8-10}
          &  $\rho$     & M (J)  & Bias (in  \%)  & RMSE   & Coverage &       & Bias (in   \%)  & RMSE   & Coverage \\
\hline    Norm  &  -0.1 & 50   &  8.01E-02 & 0.0359 & 91.94\% &       & 3.17E-02 & 0.0304 & 92.52\% \\
          &       & 100       & 3.43E-02 & 0.0483 & 89.47\% &       & 7.99E-02 & 0.0404 & 91.99\% \\
          &       & 200           & -1.93E-01 & 0.0481 & 92.18\% &       & -1.00E-01 & 0.0396 & 94.40\% \\
          &  0.2  & 50            & -8.21E-02 & 0.0258 & 91.66\% &       & -1.01E-01 & 0.0217 & 93.70\% \\
          &       & 100           & 1.31E-01 & 0.0348 & 89.87\% &       & 1.03E-01 & 0.0296 & 91.20\% \\
          &       & 200           & -1.32E-01 & 0.0343 & 92.50\% &       & -1.21E-01 & 0.0297 & 92.60\% \\
          &  0.5  & 50           & -6.38E-02 & 0.0175 & 94.12\% &       & -2.01E-02 & 0.0151 & 95.70\% \\
          &       & 100           & -2.78E-02 & 0.0240 & 92.10\% &       & -5.44E-02 & 0.0200 & 93.00\% \\
          &       & 200          & 6.24E-02 & 0.0239 & 94.30\% &       & -7.06E-03 & 0.0209 & 94.10\% \\
    LN & -0.1  & 50            & -5.01E-02 & 0.0346 & 91.99\% &       & 1.69E-02 & 0.0354 & 92.29\% \\
          &       & 100           & 7.77E-02 & 0.0478 & 89.21\% &       & 5.27E-02 & 0.0488 & 89.38\% \\
          &       & 200          & -1.38E-01 & 0.0493 & 91.98\% &       & -8.07E-04 & 0.0464 & 92.59\% \\
          &    0.2  & 50            & -5.86E-02 & 0.0261 & 91.82\% &       & -4.46E-02 & 0.0250 & 91.42\% \\
          &       & 100           & 7.04E-02 & 0.0339 & 90.16\% &       & 7.59E-02 & 0.0352 & 89.39\% \\
          &       & 200           & 3.88E-02 & 0.0340 & 92.40\% &       & 4.38E-02 & 0.0345 & 92.70\% \\
          &   0.5  & 50            & -5.39E-02 & 0.0169 & 94.43\% &       & -3.60E-02 & 0.0172 & 93.93\% \\
          &       & 100          & -1.02E-01 & 0.0241 & 93.00\% &       & -8.00E-02 & 0.0234 & 93.20\% \\
          &       & 200           & -4.62E-02 & 0.0239 & 94.40\% &       & -5.02E-02 & 0.0243 & 93.80\% \\
\hline
    \end{tabular}

   Norm denotes the normal distribution; LN denotes the lognormal distribution.
  \label{tb:mmd}
\end{table}
\begin{table} [!htbp] \centering
 \caption{Bias and RMSE of the proposed, parametric, and semiparametric methods} \begin{tabular}{cccccccccccccccccc}
 \hline     &       &       & \multicolumn{ 2}{c}{Proposed Method} &       & \multicolumn{ 2}{c}{Semiparametric Method} &       & \multicolumn{ 2}{c}{Parametric Method} \\
      \hline
      &   $\rho$ & M(J)  & Bias  & RMSE  &       & Bias  & RMSE  &       & Bias  & RMSE \\\hline
Norm & -0.1  & 50    & -0.0140 & 0.0329 &       & -0.0123 & 0.0318 &       & -0.0131 & 0.0326 \\
      &       & 100   & -0.0126 & 0.0251 &       & -0.0144 & 0.0249 &       & -0.0138 & 0.0246 \\
      &       & 200   & -0.0136 & 0.0202 &       & -0.0132 & 0.0203 &       & -0.0135 & 0.0198 \\
      & 0.2   & 50    & -0.0149 & 0.0247 &       & -0.0155 & 0.0440 &       & -0.0117 & 0.0423 \\
      &       & 100   & -0.0150 & 0.0331 &       & -0.0139 & 0.0327 &       & -0.0125 & 0.0317 \\
      &       & 200   & -0.0140 & 0.0451 &       & -0.0147 & 0.0262 &       & -0.0136 & 0.0241 \\
      & 0.5   & 50    & -0.0133 & 0.0455 &       & -0.0153 & 0.0456 &       & -0.0168 & 0.0446 \\
      &       & 100   & -0.0132 & 0.0252 &       & -0.0130 & 0.0327 &       & -0.0151 & 0.0330 \\
      &       & 200   & -0.0132 & 0.0333 &       & -0.0139 & 0.0258 &       & -0.0121 & 0.0239 \\
LN & -0.1  & 50    & -0.0152 & -0.0158 &       & -0.0122 & 0.0360 &       & 0.0689 & 0.0779 \\
      &       & 100   & -0.0131 & -0.0129 &       & -0.0120 & 0.0265 &       & 0.0758 & 0.0814 \\
      &       & 200   & -0.0131 & -0.0145 &       & -0.0127 & 0.0203 &       & 0.0799 & 0.0833 \\
      & 0.2   & 50    & -0.0158 & 0.0446 &       & -0.0139 & 0.0499 &       & 0.0706 & 0.0817 \\
      &       & 100   & -0.0120 & 0.0232 &       & -0.0141 & 0.0351 &       & 0.0754 & 0.0810 \\
      &       & 200   & -0.0136 & 0.0327 &       & -0.0129 & 0.0249 &       & 0.0807 & 0.0846 \\
      & 0.5   & 50    & -0.0158 & 0.0460 &       & -0.0156 & 0.0498 &       & 0.0705 & 0.0838 \\
      &       & 100   & -0.0129 & 0.0255 &       & -0.0120 & 0.0344 &       & 0.0791 & 0.0877 \\
      &       & 200   & -0.0145 & 0.0343 &       & -0.0134 & 0.0256 &       & 0.0826 & 0.0884 \\\hline
\end{tabular}

   Norm denotes the normal distribution; LN denotes the lognormal distribution.
  \label{tb:mmdc}
\end{table}
\begin{table}[!htbp]
  \centering
  \caption{Simulated powers for comparing tests}
    \begin{tabular}{cccccccccccccccccccccc}
\hline
   \begin{tabular}{rrrrrrr}
\multicolumn{ 7}{c}{AUC}                              \\\hline
      &       & \multicolumn{ 2}{c}{Equal Weight} &       & \multicolumn{ 2}{c}{Optimal Weight} \\
      \cline{3-4}\cline{6-7}
$\rho$ &       & M=J=50 & 100   &       & 50    & 100 \\
\hline
-0.1  &       & 0.507 & 0.741 &       & 0.723 & 0.932 \\
0.2   &       & 0.335 & 0.541 &       & 0.659 & 0.909 \\
0.5   &       & 0.327 & 0.538 &       & 0.703 & 0.936 \\
\hline
      &       &       &       &       &       &  \\
\multicolumn{ 7}{c}{pAUC}                      \\\hline
      &       & \multicolumn{ 2}{c}{Equal Weight} &       & \multicolumn{ 2}{c}{Optimal Weight} \\
          \cline{3-4}\cline{6-7}
      &       & M=J=50 & 100   &       & 50    & 100 \\ \hline
-0.1  &       & 0.156 & 0.290  &       & 0.316 & 0.599 \\
0.2   &       & 0.141 & 0.212 &       & 0.280  & 0.584 \\
0.5   &       & 0.133 & 0.187 &       & 0.266 & 0.643 \\
\hline
\end{tabular}
    \end{tabular}
  \label{tb:optip}
\end{table}
\begin{table}[!htb]
  \caption{Bias, RMSE and  coverage for simulated correlated  data}{
    \begin{tabular}{ccccccccccccccccccc}
  \hline
          &       &       & \multicolumn{ 3}{c}{AUC} &       & \multicolumn{ 3}{c}{pAUC} \\
 \cline{4-6} \cline{8-10}
          & M (J) &       & Bias (in  \%)  & RMSE   & Coverage &       & Bias (in  \%)  & RMSE   & Coverage \\
        \hline
    Norm  & 50    &       & -0.1182 & 1.0266 & 97.40\% &       & 0.0627 & 0.0184 & 97.40\% \\
          & 100   &       & 0.0302 & 2.1682 & 96.60\% &       & 0.0931 & 0.0128 & 96.60\% \\
          & 200   &       & 0.0038 & 1.5226 & 95.80\% &       & 0.0116 & 0.0090 & 96.00\% \\
    LN & 50    &       & -0.0768 & 0.0143 & 97.10\% &       & 0.0097 & 0.0125 & 97.10\% \\
          & 100   &       & -0.1126 & 0.0218 & 96.20\% &       & 0.0521 & 0.0093 & 96.80\% \\
                    & 200   &       & -0.0445 & 0.0109 & 94.90\% &       & 0.0317 & 0.0188 & 95.00\% \\
 \hline
    \end{tabular}}
  \label{tb:lbd}
  
  Norm denotes the normal distribution; LN denotes the lognormal distribution.
\end{table}

\end{document}